\documentclass[12pt]{article}

\usepackage{graphicx}

\newcommand{\nn}{\nonumber}
\newcommand{\be}{\begin{equation}}
\newcommand{\ee}{\end{equation}}
\newcommand{\ba}{\begin{eqnarray}}
\newcommand{\ea}{\end{eqnarray}}
\newcommand{\req}[1]{(\ref{#1})}
\def\={\,=\,}
\newcommand{\ci}[1]{\cite{#1}}

\def\mev{~{\rm MeV}}
\def\gev{~{\rm GeV}}

\def\als{\alpha_{\rm s}}

\newcommand{\tw}{\textwidth}

\def\vb0{{\bf b}_0}

\newcommand{\da}{{distribution amplitude}}

\def\={\,=\,}

\begin{document} 
\thispagestyle{empty}
\begin{flushright}
WU B 19-2 \\
March, 14  2019 \\[20mm]
\end{flushright}

\begin{center}
{\Large\bf On some implications of the BaBar data on the $\gamma^*\eta'$ transition form factor} \\
\vskip 10mm

P.\ Kroll \footnote{Email:  kroll@physik.uni-wuppertal.de}
\\[1em]
{\small {\it Fachbereich Physik, Universit\"at Wuppertal, D-42097 Wuppertal,
Germany}}\\
\vskip 5mm

K.\ Passek-Kumeri\v{c}ki 
\footnote{Email: passek@irb.hr}\\[1em]
{\small \it Theoretical Physics Division, Rudjer Bo\v{s}kovi\'{c} Institute, 
Zagreb, Croatia}\\

\end{center}
\vskip 5mm 
\begin{abstract}
At leading-twist accuracy the form factors for the transitions from a virtual photon to
the $\eta$ or $\eta'$ can be expanded into a power series of the variable $\omega$, being related to
the difference of two photon virtualities. The series possess the remarkable feature that
only the Gegenbauer coefficients of the meson \da s of order $l\leq m$ contribute to the 
term $\sim \omega^m$. Thus, for $\omega\to 0$ only the asymptotic meson \da{} contributes,
allowing for a test of the mixing of the $\eta$ and $\eta'$ decay constants. Employing
the Gegenbauer coefficients determined in analysis of the form factors in the real
photon limit, we present predictions for the $\gamma^*\eta$ and $\gamma^*\eta'$ form factors
and compare them to the BaBar data.   
\end{abstract}

\noindent Keywords: hard exclusive processes, perturbative QCD, meson transition form factors, eta mesons

{\it 1. Introduction} \quad The photon-meson transition form factors have always found 
much attention; there is a rich literature 
about these simple observables. These form factors have been measured in a rather large 
range of photon virtualities and the data are analyzed within the 
framework of collinear factorization. An interesting generalization of these
observables are the form factors for the transitions from a virtual photon to a meson. 
Also these form factors have repeatedly been studied
theoretically. Recently, the BaBar collaboration has measured such a form factor 
for the first time \ci{babar18}, namely the $\gamma^*\eta'$ one. 
Although the data are not very accurate this measurement is important since 
it demonstrates the feasibility of measuring such form factors at large photon virtualities. 
The prospects of getting better and more data from future experiments, as for instance 
BELLE 2, are high. Ji and Vladimirov \ci{ji19} already analyzed the BaBar data within 
the  collinear factorization approach. The authors showed agreement of the data
with perturbative QCD and put the emphasize on special features like power
corrections and effects of the binning of the data. They also elaborated on
the kinematic regions that are particularly sensitive to the underlying dynamics. 
A particular aspect of the theoretical description of the transition form factors 
are not investigated in \ci{ji19} although the authors are aware of it: the 
separation of Gegenbauer coefficients in dependence of the difference between 
the two photon virtualities, $Q_1^2=-q_1^2$ and $Q_2^2=-q_2^2$ (with the $q_i$ being the 
momenta of the photons), or rather in dependence on the variable
\be
\omega \= \frac{Q_1^2-Q_2^2}{Q_1^2+Q_2^2}\,.
\label{eq:omega}
\ee
This property of the collinear factorization approach  has first been pointed out 
in \ci{DKV1}. The purpose of the present paper is to study this property in some detail
and to generalize it to next-to-leading order (NLO) of perturbative QCD for the case 
of the $\gamma^* \eta$ and $\gamma^* \eta'$. A comparison with the BaBar data will also be made.\\

\begin{figure}
\centering
\includegraphics[width=0.45\tw]{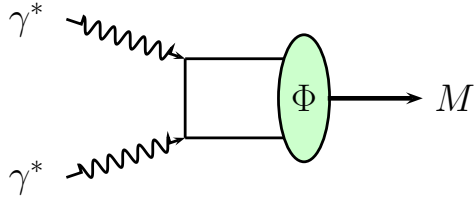}
\caption{A LO Feynman graph for $\gamma^*\;\gamma^*\to M$.}
\label{fig:graph}
\end{figure}

{\it 2. The general idea:}\quad Consider the process $\gamma^*\; \gamma^*\to M$
where $M$ is an unflavored, charge-parity even meson. The behavior of the transition form 
factors appearing in that process is, at large photon virtualities, determined by the 
expansion of a product of two electromagnetic currents about light-like distances 
\ci{brodsky-lepage}. The form factors then factorize into a hard scattering amplitude, $T_H$, 
and a soft meson matrix element, parametrized as a process-independent meson distribution 
amplitude, $\Phi_M(\xi)$, where $\xi=2x-1$ and $x$ is the usual momentum fraction carried by the 
quark inside the meson. We assume that the \da{} possesses a Gegenbauer expansion
\be
\Phi_M(\xi,\mu_F)\= (1-\xi^2)^{\lambda -1/2} \sum a_{Mn}(\mu_F) C_n^{(\lambda)}(\xi)        
\ee 
where $C_n^{(\lambda)}$ is the n-th Gegenbauer polynomial of order $\lambda$. In general
several Fock components of a meson may contribute to a particular form factor; the Gegenbauer
expansions of the corresponding \da s may have different values of $\lambda$. The Gegenbauer 
coefficients, $a_{Mn}$, depend on the factorization scale, $\mu_F$, at which the \da{} is probed. 
The hard scattering amplitude, $T_H$, has a very simple structure with the quark propagator
$\sim 1/(1\pm \xi\omega)$, see the LO Feynman graph shown in Fig.\ \ref{fig:graph}.
At leading-twist accuracy and for $\omega<1$ the hard scattering amplitude can be expanded into 
a double power series $\sum_m\omega^m p(\xi)$. Here, $p$ is a polynomial of $\xi$ of order
$m'=m+l'$ where $l'$ is a small integer, typically $|l'|\leq 1$. Consider a term 
$\omega^m \xi^k$ ($m, k$ positive integers) in that expansion. Its convolution with a \da{} reads
\be
I_{mk}\= \int_{-1}^1 d\xi (1-\xi^2)^{\lambda-1/2} \sum_n a_n C_n^{(\lambda)}(\xi)\omega^m\xi^k\,.
\ee
Any power of $\xi$ can be expressed in terms of the Gegenbauer polynomials \ci{kim12} 
\be
\xi^k \= \sum^k_{l, k-l\equiv\, 0\, mod\, 2} d^{\lambda}_{kl} C^{\lambda}_l(\xi)
\ee
where
\be
d^\lambda_{kl}\=\frac{(l+\lambda) k! \Gamma(\lambda)}
                {2^k\left(\frac{k-l}{2}\right)!\Gamma(\frac{k+l+2\lambda+2}{2})}\,.
\ee    
Using this property and applying the orthogonality of the Gegenbauer polynomials, one arrives at
\be
I_{mk}\=\omega^m \sum_n a_n \sum^k_{l, k-l\equiv\, 0\, mod\, 2} d^{\lambda}_{kl} 
                      \frac{\pi 2^{1-2\lambda} \Gamma(l+2\lambda)}
                           {l! (l+\lambda) [\Gamma(\lambda)]^2}\; \delta_{nl}\,.
\ee
Obviously, Gegenbauer coefficients $a_n$  with $n>k$ do not contribute to the integral $I_{mk}$.
This is the observation made in \ci{DKV1} for the $\gamma^*\pi^0$ form factor to NLO (the
latter corrections were taken from \ci{chase81}) and, to LO, for the $\gamma^*\eta$ and $\gamma^*\eta'$
ones. Below we are going to generalize the latter case to NLO. In \ci{melic} it has been shown
that, for the $\gamma^*\pi^0$ form factor, this property of the $\omega$-expansion even holds 
to NNLO~\footnote{
        The result generalizes to higher order of perturbative QCD provided no terms $\sim \ln{\xi}$ 
        or $\sim \ln{(1-\xi)}$ occur.}.
The correlation between the power of $\omega$ and the Gegenbauer coefficients also holds for the 
$\gamma^* f_0(980)$ \ci{kroll17} to LO and, exploiting the NLO corrections given in \ci{chase81},
also to that order. For the axial-vector, e.g.\ $\gamma^*a_1(1260)$ \ci{kopp}, and tensor, 
e.g.\ $\gamma^* f_2(1270)$ \ci{kopp,braun16}, form factors the correlation holds to LO; 
the NLO corrections are unknown as yet.  Meson-mass corrections, taken into 
account for instance in \ci{kopp,braun16}, do not spoil this property of the $\omega$ expansion 
provided $\bar{Q}^2$ is much larger than the meson mass. Under the same premise the $\eta_c$ form 
factor is another example. As shown  in \ci{KP02} such an expansion  holds for the vertex function 
of the annihilation of two virtual gluons into a pseudoscalar meson too. Results presented in \ci{ali}
are in agreement with the findings in \ci{KP02}. We anticipate similar properties for other mesons.
It should be stressed that for $\omega\to 1$, i.e.\ in the real photon limit, the hard scattering 
amplitude cannot be expanded this way and the sum of all Gegenbauer coefficients controls the transition 
form factors. However, if power corrections to the leading-twist result, accumulated in the soft 
end-point regions $\xi\to -1,1$, are taken into account the higher Gegenbauer coefficients are 
gradually suppressed \ci{piga11,agaev11,agaev14}. \\           

{\it 3. The $\gamma^*\eta$ and $\gamma^*\eta'$ form factors:}\quad
Because of $\eta - \eta'$ mixing the $\gamma^*\eta$ and $\gamma^*\eta'$ form factors
are much more complicated than the $\gamma^*\pi^{0}$ one. Even more so, there is an
additional complication at NLO due to contributions from the gluon-gluon Fock component of
the $\eta$ and $\eta'$ mesons. Thus, we have to take into account three \da s for each of the \da s:
\ba
\Phi_{Pi}(\xi,\mu_F) &=& \frac32 (1-\xi^2)\,\Big[1 +\sum_{n=2,4,\ldots} a_{Pn}^i(\mu_F)\, 
                               C^{(3/2)}_n(\xi)\Big]\,,  \nn\\
\Phi_{Pg}(\xi,\mu_F)&=& \frac{15}{8} (1-\xi^2)^2\,\sum_{n=2,4,\ldots} a_{Pn}^g(\mu_F)\,
                                     C^{(5/2)}_{n-1}(\xi)
\label{eq:da}
\ea
where $P=\eta, \eta'$ and $i=1,8$ refers to the flavor singlet and octet contributions~\footnote{
             As compared to previous work \ci{KP02,KP12} we changed the definition of the
             gluon \da{} by a factor  of 30 in order to facilitate comparison with other work.}. 
The full $\gamma^*P$ transition form factor is the sum of the 
flavor octet and singlet contributions
\be
F_{P\gamma^*}(\bar{Q}^2,\omega)\=F^8_{P\gamma^*}(\bar{Q}^2,\omega) 
                                          + F^1_{P\gamma^*}(\bar{Q}^2,\omega)
\label{eq:ff}
\ee
where 
\be
\bar{Q}^2\= \frac12\,\big(Q_1^2 + Q_2^2\big)\,.
\ee
The two parts of the form factor read
\ba
F_{P\gamma^*}^8&=&\frac1{3\sqrt{6}}\frac{f_P^8}{\bar{Q}^2}\;\int_{-1}^1 d\xi \Phi_{P8}(\xi,\mu_F)
         \frac{1}{1-\xi^2\omega^2}\Big[1 + \frac{\als(\mu_R)}{\pi}\,{\cal K}(\omega,\xi,\bar{Q}/\mu_F)\Big]\,,
                                                    \nn\\
F_{P\gamma^*}^1&=&\frac2{3\sqrt{3}}\frac{f_P^1}{\bar{Q}^2}\;\left\{\;\int_{-1}^1 d\xi 
                      \Phi_{P1}(\xi,\mu_F)
          \frac{1}{1-\xi^2\omega^2}\Big[1 + \frac{\als(\mu_R)}{\pi}\,{\cal K}(\omega,\xi,\bar{Q}/\mu_F)\Big]
                                     \right. \nn\\
            &&\left.\hspace*{0.15\tw} +\, \frac{\als(\mu_R)}{\pi}\,
             \int_{-1}^1 d\xi \Phi_{Pg}(\xi,\mu_F){\cal K}_{gg}(\omega,\xi,\bar{Q}/\mu_F) \right\}
\label{eq:result}
\ea
where $f_P^i$ is the constant of the decay of the meson $P$ trough the action of either a singlet
or octet axial-vector current. One notices that the LO term in \req{eq:result} has the simple 
expansion $\sim\sum_{n=0,2,\ldots}(\xi\omega)^n$. The NLO leading-twist results, evaluated in the 
$\overline{MS}$ scheme, for the octet and the quark part of the singlet contributions to the 
$\gamma^*P$ transition form factors, ${\cal K}$, are the same as for the $\gamma^*\pi^0$ one 
and can be taken from \ci{DKV1}.
The hard scattering amplitude for the gluon contribution which contributes to NLO, can be adapted 
from double DVCS \ci{mankiewicz,belitsky,ji-osborne}. The result is~\footnote{ 
     In \ci{agaev14}  a part of the gluonic hard scattering amplitude is given. A part corresponding
     to $\xi\to -\xi$ is lacking and when it is added agreement with our results is to be seen.} 
\ba
{\cal K}_{gg}&=& 
\frac1{3(1-\xi^2)\omega^2}\;
\left\{
-\frac{4(1-\omega)}{1-\xi}\ln{(1-\omega)} 
+\frac{1-\omega}{1-\xi}\ln^2{(1-\omega)} 
\right.  \nn\\ & &  \left.
+ \frac{4-\omega(1+\xi)^2}{1-\xi}\ln^2{(1-\xi\omega)} 
- \frac{2-\omega(1+\xi)}{2(1-\xi)}\ln^2{(1-\xi\omega)} 
\right. \nn\\& &  \left.
+ 
\frac{1-\omega}{1-\xi}
               \Big(2\ln{(1-\omega)} - (1+\xi)\ln{(1-\xi\omega)}\Big)
 \ln{\frac{\bar{Q}^2}{\mu_F^2}}
\right\}  
\nn\\
   && + (\omega\to -\omega) - (\xi\to -\xi) - (\xi\to -\xi, \omega\to -\omega) 
\nn\\[0.2cm]
      &=& 
-\frac{5}{9}\xi\,\omega^2\,\Big(1-\frac25\ln{\frac{\bar{Q}^2}{\mu_F^2}}\Big) 
-\frac{37}{135}\xi (1+2\xi^2)\,\omega^4\,\Big(1-\frac{12}{37}\ln{\frac{\bar{Q}^2}{\mu_F^2}}\Big) 
          + {\cal O}(\omega^6)\,.  
\nn\\ &&   
\label{eq:gg}
\ea
As one sees this is an expansion of the type discussed in Sect. 2.
Power corrections to the above leading-twist result are mainly accumulated in the soft end-point
regions where $\xi\to \pm 1$. They are expected to be small for small $\omega$ and large $\bar{Q}^2$.
This is obvious from the parton propagator $1/(1 - \xi^2\omega^2)$ in \req{eq:result}. For $\omega\to 0$
the form factor becomes less sensitive to the end-point regions. Estimates of power corrections
arising from quark transverse momentum \ci{DKV1} or from meson-mass corrections \ci{agaev14} support
this expectation. Thus, for the case of interest, it seems to be reasonable to work at leading-twist
accuracy. \\

{\it 4. Expansion of the NLO leading-twist $\gamma^*\eta$ and $\gamma^*\eta'$ form factors:}\quad
Expanding the form factor \req{eq:ff} upon $\omega$ leads to
\be
F_{P\gamma^*}(\bar{Q}^2,\omega)\=\frac{\sqrt{2}}{3\bar{Q}^2}\;\sum_{n=0,2,\ldots} 
                                 c_{Pn}(\bar{Q}^2)\,\omega^n
\label{eq:FF-expansion}
\ee
where the first coefficients of the series read
\ba
c_{P0}&=&f_P\,\Big(1-\frac{\als}{\pi}\Big)\,, \nn\\
c_{P2}&=&\frac{f_P}{5}\;\Big(1-\frac53\frac{\als}{\pi}\Big)
                                  +  \frac{12}{35}\;a_{P2}^{\rm eff}\;
                                      \Big(1+\frac5{12}\frac{\als}{\pi}\,
                             \Big(1-\frac{10}{3}\ln{\frac{\bar{Q}^2}{\mu_F^2}}\Big)\Big)\nn\\
   &-& \sqrt{\frac23}\;\frac{50}{63}\; f_P^1 a_{P2}^g \;\frac{\als}{\pi}\,
              \Big(1-\frac25\ln{\frac{\bar{Q}^2}{\mu_F^2}}\Big)\,,   \nn\\
c_{P4}&=& \frac{3f_P}{35}\;\Big(1-\frac{59}{27}\frac{\als}{\pi}\Big)  
                + \frac{8}{35}\; a_{P2}^{\rm eff}\;
                  \Big(1+\frac{173}{216}\;\frac{\als}{\pi}\;
                        \Big(1-\frac{300}{173}\ln{\frac{\bar{Q}^2}{\mu_F^2}} \Big)\Big)\nn\\
     &+&  \frac{8}{77}\; a_{P4}^{\rm eff}\;
                                \Big(1+\frac{523}{270}\;\frac{\als}{\pi}\;
                      \Big(1-\frac{546}{523}\ln{\frac{\bar{Q}^2}{\mu_F^2}} \Big)\Big)\nn\\ 
   &-&  \sqrt{\frac23}\;\frac{370}{567}\;\frac{\als}{\pi}\;
                        \Big(1-\frac{12}{37}\ln{\frac{\bar{Q}^2}{\mu_F^2}}\Big)\;
                         f_P^1 \; \Big(a_{P2}^g+\frac{28}{55}a_{P4}^g\Big)\,. 
\label{eq:cn}
\ea
The effective decay constant is defined by
\be
f_P\= \frac1{\sqrt{3}}\,\Big[f_P^{8} + 2\sqrt{2} f_P^{1}\Big]\,, 
\ee
and the effective quark Gegenbauer coefficients by
\be
a^{\rm eff}_{Pn}(\mu_F)\=\frac1{\sqrt{3}}\,\Big[f_P^{8} a^8_{Pn}(\mu_F)
                         +2\sqrt{2}f_P^{1} a^1_{Pn}(\mu_F)\Big]\,.  
\ee
The various Gegenbauer coefficients depend on the factorization scale, $\mu_F$. Thus,
\ba
a_{Pn}^8(\mu_F)&=& a_{Pn}^8(\mu_0)\,L^{\gamma_n^{qq}/\beta_0}\,,  \nn\\
a_{Pn}^1(\mu_F)&=& \frac1{1-\rho_n^{(+)}\rho_n^{(-)}}\,\left[\Big(L^{\gamma_n^{(+)}/\beta_0}
                           -\rho_n^{(+)}\rho_n^{(-)}L^{\gamma_n^{(-)}/\beta_0}\Big)\,a_{Pn}^1 (\mu_0^2)\right.\nn\\
         &&\left. \hspace*{0.15\tw}+\, \Big(L^{\gamma_n^{(-)}/\beta_0}-L^{\gamma_n^{(+)}/\beta_0}\Big)\,
                                 \rho_n^{(-)}\,a_{Pn}^g(\mu_0^2)\right], \nn\\
a_{Pn}^g(\mu_F)&=&\frac1{1-\rho_n^{(+)}\rho_n^{(-)}}\,\left[\Big(L^{\gamma_n^{(-)}/\beta_0}
                            -\rho_n^{(+)}\rho_n^{(-)}L^{\gamma_n^{(+)}/\beta_0}\Big)
                                  \,a_{Pn}^{g}(\mu_0^2) \right. \nn\\
         &&\left.\hspace*{0.15\tw} +\, \Big(L^{\gamma_n^{(+)}/\beta_0}-L^{\gamma_n^{(-)}/\beta_0}\Big)\,
                                               \rho_n^{(+)}\,a_{Pn}^{1}(\mu_0^2)\right].
\label{eq:evolution}
\ea                    
The parameters $\rho_n^{(\pm)}$ read
\be
\rho_n^{(+)}\=\frac15\frac{\gamma_n^{gq}}{\gamma_n^{(+)}-\gamma_n^{gg}}\,, \qquad
\rho_n^{(-)}\=5\frac{\gamma_n^{qg}}{\gamma_n^{(-)}-\gamma_n^{qq}}\,.
\label{eq:rho}
\ee
and
\be
L\=\frac{\als(\mu_0)}{\als(\mu_F)}
\ee
The anomalous dimensions, $\gamma_n^i$, can, for our conventions, be found in \ci{KP12} ($\beta_0=25/3$
for four flavors)~\footnote{
     Because of the definition of the gluon \da{} in Eq.\ \req{eq:da} the quantities $\rho_n^{(\pm)}$
     differ from those quoted in \ci{KP12} by the factor of 30.}. 
As we see from \req{eq:cn} the Gegenbauer coefficients $a_{Pn}^{i(g)}$ are 
suppressed in the transition form factors by a power $\omega^n$. Thus, accurate data on the 
transition form factors, $F_{P\gamma^*}$, offer the possibility to measure at least the lowest
Gegenbauer coefficients of the meson \da s. This is to be contrasted with other hard exclusive
processes where frequently the $1/\xi$-moment of the \da s controls the observables. To this
moment all Gegenbauer coefficients contribute equally. The expansion coefficients $c_{Pn}$
depend on $\bar{Q}^2$ only logarithmically through $\als$ and the evolution. The transition
form factors $F_{P\gamma^*}$ scale as $1/\bar{Q}^2$. 

{\it 5. Comparison with the BaBar data:}\quad 
The most interesting result is that, for $\omega\to 0$, the leading term of the transition form
factor only depends on the asymptotic meson \da s:
\be
\bar{Q}^2 F_{P\gamma^*} \= \frac{\sqrt{2}}{3} f_P\,\Big(1-\frac{\als}{\pi}\Big)\; 
                                                             + {\cal O}(\omega^2,\als^2)\,.
\ee 
This result has been already given in \ci{DKV1}. In contrast to the case of the pion where the 
decay constant $f_\pi$ is known, $f_P$ depends on the $\eta - \eta'$ mixing parameters. In the 
two-angle mixing scheme \ci{FKS1} the decay constants, $f_P^i$, are parametrized as
\ba    
f_{\eta'}^8&=& f_8 \sin{\theta_8}\,, \qquad f_{\eta'}^1\=f_1 \cos{\theta_1}\,, \nn\\
f_\eta^8&=& f_8\cos{\theta_8}\,, \qquad f_{\eta}^1\=-f_1 \sin{\theta_1}\,.
\ea 
The various mixing parameters are taken from the phenomenological set presented in \ci{FKS1}
($f_\pi=131\,\mev$)
\ba
f_8&=& (1.26\pm 0.06)\,f_\pi\,,  \qquad \theta_8\=-21.2\pm 1.4\,,   \nn\\
f_1&=& (1.17\pm 0.04)\,f_\pi\,, \qquad \theta_1\=-9.2\pm 1.4\,.
\label{eq:pheno}
\ea
In contrast to $f_8$ and the mixing angles the singlet decay constant, $f_1$, is
renormalization scale dependent \ci{leutwyler}. The anomalous dimension controlling this 
scale dependence is of order $\als^2$ and therefore small. In the determination of the mixing 
parameters this scale dependence is usually ignored. It is therefore not clear at which scale 
\req{eq:pheno} holds. Since the data on the $\gamma\eta$ and $\gamma\eta'$ transition form 
factors \ci{cleo} as well as a number of charmonium decays play an important role in the 
analysis of $\eta -\eta'$ mixing \ci{FKS1} a possible initial scale of $f_1$ presumably
lies in the range of $2 - 4\,\gev^2$. If so, the scale dependence of $f_1$ is weak; its effect
on the form factors is of the order of the theoretical errors quoted in Tab.\ \ref{tab:default}. 
We therefore ignore the scale dependence of $f_1$ in the following. This procedure is 
consistent with the determination of the mixing parameters \req{eq:pheno}.    

For the evaluation of the form factors the QCD coupling, $\als$, is evaluated from the two-loop 
expression with $\Lambda_{QCD}=319\,\mev$ for four flavors in the $\overline{MS}$-scheme \ci{pdg}. 
The renormalization scale is chosen as $2\bar{Q}^2$. This is all we need for an evaluation
of the transition form factors at $\omega=0$. The results of the computation 
are presented in Tab.\ \ref{tab:default} and compared to the three BaBar data points at this 
value of $\omega$. Excellent agreement is to be observed, all predicted values agree with the 
experimental ones within the admittedly large experimental errors. The mixing parameters have 
been determined repeatedly, using more recent but often less data, e.g.\ \ci{escribano-frere,cao12}. 
Also these sets of mixing parameters agree with the BaBar data within the experimental errors. 
Even the theoretical mixing parameters quoted in \ci{kroll-isospin} which, with the help of the 
divergences of the axial-vector currents, are expressed in terms of particle masses, agree with 
experiment. The predictions for the $\gamma^*\eta'$ form factor obtained from the various sets 
of mixing parameters, even agree  within the parametric phenomenological errors quoted in Tab.\ 
\ref{tab:default}, except for the mixing parameters of \ci{escribano-frere} which lead
to values of the $\gamma^*\eta'$ form factor about $2\,\sigma$ (with respect to the phenomenological
errors) larger than the predictions listed in the table. 
A somewhat larger spread of the predictions 
for the $\gamma^*\eta$ form factor is obtained. 
\begin{table*}[t]
\renewcommand{\arraystretch}{1.4} 
\begin{center}
\begin{tabular}{| c | c || c | c | c | c | }
\hline   
$\bar{Q}^2[\gev^2]$ &  $\omega$ & $\bar{Q}^2 F_{\eta'\gamma^*}^{\rm exp}[\mev]$ 
                     & $\bar{Q}^2 F_{\eta'\gamma^*}[\mev]$ 
                     &  $\chi^2$ & $\bar{Q^2} F_{\eta\gamma^*}[\mev]$\\[0.2em]  
\hline
6.48  & 0.000 & $92.8\pm 13.8$  & $92.7\pm 3.9$ & 0.00 & $56.2\pm 3.3$ \\[0.2em]
16.85 & 0.000 & $90.1\pm 37.3$  & $93.8\pm 3.9$ & 0.01 & $56.8\pm 3.3$ \\[0.2em]
9.55  & 0.553 & $78.7\pm 13.5$  & $98.7\pm 4.1$ & 2.19  & $59.9\pm 3.5$ \\[0.2em]
26.53 & 0.436 & $161.0\pm 44.2$ & $97.7\pm 4.1$ & 2.05  & $59.2\pm 3.5$ \\[0.2em]
45.63 & 0.000 & $397.4\pm 400.9$ & $94.6\pm 4.0$ & 0.57  & $57.4\pm 3.4$ \\[0.2em]
\hline
\end{tabular}
\end{center}
\caption{Predictions for the scaled $\gamma^*\eta$ and $\gamma^*\eta'$ transition factors. 
The data are taken from \ci{babar18} and the mixing parameters from \ci{FKS1} 
(phenomenological values). For $\omega\neq 0$ the Gegenbauer coefficients \req{eq:gegenbauer2} 
are used. Parametric errors of the theoretical results are also quoted. The $\chi^2$ values 
are evaluated with regard to the experimental errors.}
\label{tab:default}
\renewcommand{\arraystretch}{1.0}   
\end{table*} 

The BaBar collaboration has also measured the $\gamma^*\eta'$ form factor for two non-zero 
but adjacent values of $\omega$. The two face values of the form factor data differ by about a 
factor of two. It seems difficult to accommodate this difference within the NLO leading-twist 
theory since, as we mentioned above, the expansion coefficients, $c_{Pn}$, only depend on 
$\bar{Q}^2$ logarithmically. Anyway the present experimental information on the 
$\omega$-dependence of the transition form factors is too limited for an attempt of fitting
even the lowest Gegenbauer coefficients of the meson \da s. Nevertheless, we can carry out the
following check: We make use of the second Gegenbauer coefficients which we extracted in \ci{KP12}
from the data on the meson-photon transition form factors \ci{cleo,babar11}. In this analysis
we have had to assume that all higher Gegenbauer coefficients do not contribute in the real-photon
limit. Thus, although the extracted Gegenbauer coefficients are to be regarded as
effective ones, perhaps contaminated by higher-order Gegenbauer coefficients, we here identify them 
with the real second order coefficients. It should also be mentioned that in \ci{KP12} 
particle-independence of the meson \da s is assumed, i.e.\
\be
\Phi^{i(g)}_P \=\Phi^{i(g)}
\ee 
This plausible assumption has also been made in \ci{agaev14,FKS1}.
The values of the Gegenbauer coefficients read \ci{KP12}
\be   
 a_2^8\=-0.05\pm 0.02\,, \quad a_2^1\=-0.12\pm 0.01\,, \quad a_2^g\=0.63\pm 0.17\,,
\label{eq:gegenbauer2}
\ee
valid at the scale $\mu_0=1\,\gev$. In order to match the choice of the factorization scale 
made in \ci{KP12} in the real photon case we choose $\mu_F^2=2\bar{Q}^2$. From the Gegenbauer 
coefficients \req{eq:gegenbauer2} we obtain the values for the $\gamma^*\eta$ and $\gamma^*\eta'$ 
transition form factors quoted in Tab.\ \ref{tab:default}. Terms $\propto \omega^4$ are 
included in that evaluation. Both the results at the non-zero values of $\omega$ deviate
by about $2\,\sigma$ from experiment. One of the theoretical values is too low as compared
to the BaBar data, the other one too high. We emphasize - a change of the Gegenbauer 
coefficients \req{eq:gegenbauer2} either increase or decreases the values of the form factors 
for both the $\omega$ values. Thus, it seems that we cannot improve the predictions. However, 
we do not think that serious conclusions should be drawn from this result; more accurate data 
are required for this. We remark that the $\omega^2$-term affects the results by about $5\%$, 
the $\omega^4$ term ( with zero fourth-order Gegenbauer coefficients) by less than $1\%$. The 
contribution from the gluon-gluon Fock component of the mesons, $\propto a_n^g$, is tiny but, 
implicitly, it substantially affects the results through the evolution of $a_n^1$, see 
\req{eq:evolution}.

\begin{figure}
\centering
\includegraphics[width=0.45\tw]{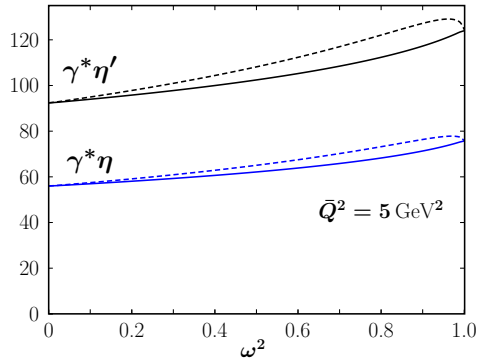}
\caption{Predictions for the  scaled $\gamma^*\eta$ and $\gamma^*\eta'$ form factors versus
$\omega^2$ at $\bar{Q}^2=5\,\gev^2$. Solid (dashed) lines: fit \req{eq:gegenbauer2}
(alternative fit - see text). }
\label{fig:FF}
\end{figure}

In Fig.\ \ref{fig:FF} we display predictions for the scaled form factors versus $\omega^2$ at 
$\bar{Q}^2=5\,\gev^2$. In this computation we have naturally used the expressions \req{eq:ff}
and \req{eq:result} for the form factors instead of the expansion \req{eq:FF-expansion}. 
We stress again - the scaled form factors depend only logarithmically on 
$\bar{Q}^2$. For comparison we have made an alternative evaluation of the transition form factors 
for which we have assumed $a_2^8=a_2^1=0.25$, positive values for these Gegenbauer coefficients are 
favored by QCD sum rules \ci{agaev14}, and, in order to have the same effective $a_2^i$ values in 
the real photon limit, we have chosen the following (effective) fourth order Gegenbauer coefficient: 
$a_4^8=-0.31$ and $a_4^1=-0.38$. The gluon \da{} is left unchanged. As one sees from Fig.\ \ref{fig:FF}
the two sets of Gegenbauer coefficients lead to the same form factors for $\omega\to 0$ and,
indeed, in the real photon limit. However, for large, but $<1$, values of $\omega$ the predictions
differ and for sufficiently accurate data on the $\gamma^*P$ form factors one may distinguish
between the two sets of Gegenbauer coefficients. Thus, we conclude data on the $\gamma^*P$ transition
form factors may provide more detailed information on the meson distribution amplitudes then
one obtains in the real photon limit.
    
{\it 6. Summary:} \quad
We have discussed the $\omega$-expansion of the form factors for the annihilation
of two virtual photons into a meson to leading-twist accuracy and have, in particular, investigated 
the correlation between the power of $\omega$ and the Gegenbauer coefficients of the 
corresponding meson \da s in some detail. We have applied this property to the $\gamma^*\eta$ and
$\gamma^*\eta'$ form factors and have shown that to the $\omega^m$-term only the Gegenbauer
coefficients of the quark octet and singlet \da s as well as of the gluon \da{} to order $n\leq m$
contribute. While for the quark \da s this property has already been discussed in \ci{DKV1}
the gluon contribution is new.

The correlation between the power of $\omega$ and the Gegenbauer coefficients 
is a possibility to learn more about the meson \da s as it is possible from the transition form factors
in the real photon limit. In the latter case the sum of all Gegenbauer coefficients makes up the form
factors. While for the $\gamma^*\pi^0$ form factor the application of the $\omega$ expansion is
very simple and straightforward it is more involved for the $\eta$ and $\eta'$ because of the $\eta-\eta'$
mixing and the contributions from the gluon-gluon Fock components of these mesons. But is is
feasible with the plausible assumption of particle-independence of the corresponding \da s. Still there 
are three Gegenbauer coefficients at any order but only two independent form factor measurements.
The gluon \da{} is only separated from the the two quarks ones through the $\als$-corrections
which merely provide a small lever arm in practice. In any case, accurate form factor data
for large $\omega$, close to 1, will certainly allow to check whether the effective value of 
$a_2$ extracted in the real  photon limit, results from the cancellation 
of rather large individual terms or from the smallness of the $a_n$ for $n>2$ as stated in \ci{DKV1}. 
The much higher accuracy of the data on $F_{P\gamma}$ than those 
for $F_{P\gamma^*}$ could thus be 
overcompensated. 

{\it Acknowledgements}
Partially supported by the European Union through the European Regional
Development Fund - the Competitiveness and Cohesion Operational Programme
(KK.01.1.1.06).

\end{document}